\documentclass[epj,final]{svjour}

\usepackage{latexsym}
\usepackage{url}
\usepackage{amsfonts}
\usepackage{amsmath, amssymb}
\RequirePackage{graphicx}

\begin{document}

\title{Hubble tension vs two flows}
%\shorttitle{Hubble tension vs two flows} %Insert here a short version of the title if it exceeds 70 characters

\author{V.G. Gurzadyan\inst{1,2} \and A. Stepanian\inst{1}}
%\shortauthor{V.G. Gurzadyan, A. Stepanian}

\institute{                    
  \inst{1} Center for Cosmology and Astrophysics, Alikhanian National Laboratory and Yerevan State University, Yerevan, Armenia\\
  \inst{2} SIA, Sapienza University of Rome, Rome, Italy
}
%\pacs{98.80.-k}{Cosmology}
%\pacs{nn.mm.xx}{Second pacs description}
%\pacs{nn.mm.xx}{Third pacs description}

\abstract{The Hubble tension is shown to be solvable, without any free parameter,  conceptually and quantitatively, within the approach of modified weak-field General Relativity involving the cosmological constant $\Lambda$. That approach enables one to describe in a unified picture both the dynamics of dark matter containing galaxies and the accelerated expansion of the Universe, thus defining a {\it local} Hubble constant of a local flow and the {\it global} one. The data on the dark matter content of peculiar galaxy samples are shown to be compatible to that unified picture. Future more refined surveys of galaxy distribution, hierarchical dynamics and flows within the vicinity of the Local group and the Virgo supercluster can be decisive in revealing the possible common nature of the dark sector.}

%\begin{document}

\maketitle

\section{Introduction}

The Hubble tension as the claimed discrepancy between the values of the Hubble parameter associated to the early and late Universe treatments, is attracting much attention \cite{AR,M106,VTR,Pe,Dh,R,AR4,BV,Pl}. The key issue is whether it is a signature of principally new physical concepts or outlines the need of more accurate observational data analysis and interpretation within the existing concepts. At the same time, the dark sector continues to remain one of key problems of cosmology and fundamental physics and its possible link with the Hubble tension is a natural quest.  

Among the models proposed to explain the available observational data on the dark matter (DM) is the one based on a modification of General Relativity (GR) with the cosmological constant $\Lambda$ entering its weak-field limit \cite{G,GS1,GS2}. That approach follows from the Newton theorem on the equivalency of sphere's gravity and that of a point mass situated in its center. Within that approach both the dark matter and dark energy are determined by cosmological constant $\Lambda$ which acts as a second fundamental constant of gravity along with $G$ and the DM is defined by weak-field limit of GR \cite{GS1,GS2,GS4,GS3}. That $\Lambda$-gravity approach enables one to explain the dynamical properties of groups and clusters of galaxies \cite{G,GS2}.   Preliminary, the $\Lambda$-gravity vs the $H$-tension was considered in \cite{GS3}, and now more refined data are available which are analysed below. These data enable one to define {\it local} and {\it global} Hubble constants, to reveal their quantitative difference over the distance ladder and to show, without any free parameter, their correspondence to the cosmological parameters. 
Along with that, we show that recently studied galaxy samples \cite{WDM1,WDM2,WDM3}, either claimed with no DM or as made up mostly ($98 \%$) of DM \cite{Dr44}, are also compatible to the $\Lambda$-gravity.

Among other approaches regarding the dynamics of the Local group surroundings we mention e.g. \cite{RG,Dav,Kar,Ch,VC,Na1,Ba,Mc,Chr} and the references therein, based on various assumptions or gravity modifications.

Within the $\Lambda$-gravity approach discussed below, the Hubble parameter defining the cosmological model (the early Universe) and those obtained at distance ladder studies (late Universe), are explained naturally, without an extra parameter, as a consequence of the common nature of dark energy and dark matter. While the observational data  indicate the robust value of the Hubble parameter for the {\it local volume} galactic system dynamics \cite{AR1,AR2,AR3,J}, we show the compatibility of $\Lambda$-gravity to flow dynamics at several scales around Local Group.    

Importantly, the Hubble tension, in view of the indications revealed in the analysis below, thus can act as an independent test for the modified weak-field General Relativity, complementing the possibilities of gravity lensing \cite{Col,GS5}, celestial mechanics 
\cite{KEK}, galaxy cluster dynamics and cosmic voids \cite{Cap2,Cap3,GKS,GK}, cosmological perturbation evolution \cite{E} and dedicated GR experimental programs  \cite{Tur,Ciu}.

\section{Newton theorem}

The Newton theorem on ``sphere-point" equivalency enables one to arrive to the weak-field modification of General Relativity given by the metric \cite{GS1} $(c=1)$
\begin {equation} \label {mod}
g_{00} = 1 - \frac{2 G m}{r} - \frac{\Lambda r^2}{3}; 
\qquad g_{rr} = \left(1 - \frac{2 G m}{r} - \frac {\Lambda r^2}{3}\right)^{-1}.
\end {equation} 

This metric was known before (Schwarzschild - de Sitter metric), however when deduced based on Newton theorem, it provides a description of astrophysical structures such as the galaxy clusters within the weak-field limit of GR \cite{GS2}.  

The general function for force $\mathbf{F}(r)$ satisfying Newton's theorem has the form (see \cite{G1,G,GS1}) 
\begin{equation}\label{FandU}
\mathbf{F}(r) = \left(-\frac{A}{r^2} + Br\right)\hat{\mathbf{r}}\ .
\end{equation}   
The second term here leads to the cosmological term in the solutions of Einstein equations and the cosmological constant $\Lambda$ appears in weak-field GR \cite{GS3}. 

The appearance of $\Lambda$s both in Eq.(\ref{mod}) and Eq.(\ref{FandU}) has a clear group-theoretical background.  Namely, depending on the sign of $\Lambda$ - positive, negative or zero - one has three different vacuum solutions for Einstein equations corresponding to isometry groups, as shown in 
Table 1.

\begin{table}
\caption{Vacuum solutions for GR}\label{tab1}
\centering
\scalebox{1}{
\begin{tabular}{|p{0.9cm}||p{2.4cm}|p{2.4cm}||}
\hline

Sign& Spacetime&Isometry group \\ \hline
\hline
$\Lambda > 0$ &de Sitter (dS) &O(1,4)\\ \hline
$ \Lambda = 0$ & Minkowski (M) & IO(1,3)\\ \hline
$\Lambda <0 $ &Anti de Sitter (AdS) &O(2,3)\\ 
\hline
\end{tabular}
}
\end{table}

These maximally symmetric Lorentzian 4D-geometries have Lorentz group O(1,3) as their isometry stabilizer group.  The group O(1,3) of orthogonal transformations in these Lorentzian geometries implies spherical symmetry (in Lorentzian sense) at each point of spacetime, so that for all these cases O(3) is the stabilizer group of spatial geometry, that is each point (in spatial geometry) admits O(3) symmetry. This statement can be regarded as group theory formulation of Newton theorem \cite{GS3}.
 
The next important fact is that, the force of Eq.(\ref{FandU}) defines non-force-free field inside a spherical shell, thus drastically contrasting with Newton's gravity when the shell has force-free field in its interior. The non-force-free field agrees with the observational indications that galactic halos do determine features of galactic disks \cite{Kr}. The weak-field GR thus ensures that any matter, seen or unseen \cite{Ge}, at large galactic scales is interacting by the law of Eq.(\ref{mod}) and for which the virial theorem yields \cite{G}
\begin{equation}   
\Lambda=\frac{3\sigma^2}{2 c^2 R^2}\simeq 3\,\, 10^{-52} (\frac{\sigma}{50\, km s^{-1}})^2(\frac{R}{300 \,kpc})^{-2} m^{-2},
\end{equation} 
where $\sigma$ is the velocity dispersion at a given radius of halo. 

This relation in \cite{GS1,GS4} was shown to be compatible with the observed dynamics of groups and clusters of galaxies, where the value of the cosmological constant was derived from the dynamics of those galactic systems, depending on their degree of virialization. 
Further analysis of the compatibility of Eqs.(\ref{mod},\ref{FandU}) to the galactic systems, involving, importantly, {\it extremal} galaxies, i.e. claimed as DM-free or DM-rich, is performed in Section 4.   

\section{Two flows with $\Lambda$: local and global}  

According to Eq.(\ref{mod}) (for details see \cite{GS3}) the same cosmological constant enters both in the FLRW cosmological equations (as dark energy) and in the weak-field GR (as dark matter) to define galactic system structure and dynamics. Then, one has two different flows and two different Hubble constants respectively
\begin{equation}\label{Hl}
H_{local}^2 = \frac{8 \pi G \rho_{local}}{3} + \frac{\Lambda c^2}{3},
\end{equation}
\begin{equation}\label{Hg}
H_{global}^2 = \frac{8 \pi G \rho_{global}}{3} + \frac{\Lambda c^2}{3} - \frac{k c^2}{a^2(t)},
\end{equation}
where $k$ is the spatial curvature of FLRW metric and $a(t)$ is the scale factor.  Hence,  \textit{local} and \textit{global} values of Hubble constant do arise. Namely, while in Eq.(\ref{Hl}) we are dealing with the density of local universe, the measurement of $H_{global}$ in Eq.(\ref{Hg}) becomes related to cosmological parameters and the geometric features of FLRW metric. 

Then, for the $\Lambda$-gravity it is possible to obtain a flow in the local universe. Namely, one can define a \textit{critical radius} where the $\Lambda$-term in Eq.(\ref{mod}) becomes dominant
\begin{equation}\label{rcrit}
r_{crit}^3 = \frac{3 GM}{\Lambda c^2}. 
\end{equation}
Accordingly, due to repulsive nature of $\Lambda$-term, one can conclude that beyond $r_{crit}$ the gravitational field of the central object becomes repulsive and that can cause the local H-flow.
Meantime, for $\Lambda$-gravity we find one more limit, besides the Newtonian limit (related to the Newtonian term in Eq.(\ref{mod})), beyond which the second term's contribution becomes important
\begin{equation}\label{rel}
\frac{\Lambda r^2}{3} << 1, \quad r\approx 5.33 \quad Gpc.
\end{equation} 

It should be noticed that Eq.(\ref{Hl}) should not be confused as the non-relativistic limit of FLRW equations. Indeed, both from conceptual and fundamental points of view Eq.(\ref{Hl}) has nothing to do with the FLRW universe. It is obtained based on the McCrea-Milne model \cite{MM1,MM2} and the consideration of $\Lambda$-gravity. Namely, we get the Eq.(\ref{Hl}) since in the equations of McCrea-Milne model we use the gravitational potential energy according to $\Lambda$-gravity i.e.
\begin{equation}
\Phi = -\frac{GMm}{r} - \frac{\Lambda c^2 r^2 m}{6}.
\end{equation}
In this sense, the local H-flow occurs in all non-relativistic limits due to the presence of $\Lambda$ in the weak-field limit and not as a result of residuals of the expansion of the Universe. Speaking in other words considering the $\Lambda$-gravity, we will have the gravitational repulsion beyond the $r_{crit}$ and as a result of that, no matter our relativistic background geometry is dynamic, static or even collapsing   we will get a local H-flow according to Eq.(\ref{Hl}).

The observations indicate that the Universe is flat, i.e. $k=0$ in Eq.(\ref{Hg}). Thus, although Eq.(\ref{Hl}) and Eq.(\ref{Hg}) are similar in their form, as mentioned above, their essense is rather different.
The two currently indicative values of the Hubble constant are those determined by $Planck$ \cite{Pl} and $HST$ \cite{AR} data, namely
\begin{equation}\label{H}
Planck : H_{global}=67.4 \pm 0.5 \quad  kms^{-1}Mpc^{-1}
\end{equation}
\begin{equation}
HST : H_{local}= 74.03 \pm 1.42 \quad kms^{-1}Mpc^{-1}.
\end{equation}

Currently, after the publication of \cite{GS3}, new results on accurate measurements of $H_0$ have been reported.
In \cite{F}, the authors have obtained $H_0 = 69.8 \pm 1.9$ $kms^{-1}Mpc^{-1}$ . The importance of their reported value lies on the fact that they have analyzed the data of $HST$ which has been generally used to obtain the ``local $H$". However, according to authors they have used the calibration of the Tip of the Red Giant Branch (TRGB) applied to Type Ia supernovae (SNe Ia) to measure the $H$. Considering the nature of measurement and the fact that SNe Ia is used to make measurements in $0.03<z<0.4$ redshifts which is equivalent to scales (in $Mpc$) about $132 < r < 1597$, we have to conclude that the reported value should be closer to ``global $H$" in \cite{Pl}. Again, while the nature of the global flow is related to the FLRW equations, the local flow i.e. the recession of galaxies from the center of a gravitationally bound object would occur due to the presence of $\Lambda$ in the weak-field limit of GR. Since the mentioned distance scales are far larger than of any gravitationally bound structure's, it is expected to have a value for $H$ closer to "global" rather than the "local" one. 

Let us mention also the recent paper by  H0LiCOW team \cite{Wong}, where the reported value of Hubble parameter is closer to the local $H$ and is in tension with the global flow. This case drastically differs from the above mentioned one, since it is based on the gravitational lensing which itself is a local effect. As a result, the time delay in \cite{Wong} is related to “time-delay distance” which itself is sensitive to $H$ (nevertheless, the authors have mentioned that "although there is a weak dependence on  other  parameters" ), and hence does not give an independent information about the "global" structure of spacetime geometry; for details see the analysis of lensing effect for modified gravity \cite{GS5}.

The next remarkable reported data are of \cite{M106}, where the value  $H= 73.5 \pm 1.4$ $kms^{-1}Mpc^{-1}$ is obtained, via measuring the distance to M106 galaxy using its supermassive black hole. In this case, considering the nature of measurement, the reported value should be in agreement with $H_{local}$ in Eq.(\ref{H}). Note, that another analysis related to the measurement of ``local H" is \cite{Chen}, with reported values of $H$ (in $kms^{-1}Mpc^{-1}$)  shown in Table{\ref{tab1}}, which are in agreement with the value of $H$ in \cite{M106}.

Thus, within our approach the discrepancy between the reported values of $H$ is a result of measuring of two parameters defining two different dynamical effects, i.e. \textit{local} and \textit{global} H-flows. Quantitatively, as follows from Eq.(\ref{Hl}) and Eq.(\ref{Hg}), that discrepancy is related to the discrepancy between the definitions of the local and global densities i.e. $\rho_{local}$ and $\rho_{global} = 8.5 \times 10^{-27}\, Kg\, m^{-3}$ \cite{Pl}.

By considering Eq.(\ref{Hl}) and the reported values of $H$ \cite{M106,AR1,AR2,AR3,J} we obtain the local density $\rho_{local}$. Then, for three hierarchical systems we find the distances (in $Mpc$) with respect to the central object where the local H-flow occurs. Our analysis shows that, on the one hand, the results are in an exact agreement with observations and, on the other hand, with the theoretical principles. Finally, we also obtain an estimation of mass for Laniakea Supercluster.

Namely, according to Eq.(\ref{Hl}) the estimations for $\rho_{local}$ yield (in $Kg$ $m^{-3}$):
\begin{equation}\label{densloc}
\rho_{local} = 4.231 ^{0.391}_{0.384} \times 10 ^{-27}\, \quad \textrm{\cite{M106}};
\end{equation}
\begin{equation}
\rho_{local} = 4.15 ^{0.48}_{0.47} \times 10 ^{-27}\, \quad \quad \textrm{\cite{AR1}};
\end{equation}
\begin{equation}
\rho_{local} = 4.237 ^{0.454}_{0.444} \times 10 ^{-27}\, \quad \textrm{\cite{AR2}};
\end{equation}
\begin{equation}
\rho_{local} = 4.21 ^{0.46}_{0.45} \times 10 ^{-27}\, \quad \textrm{\cite{AR3}};
\end{equation}
\begin{equation}
\rho_{local} = 4.87 ^{1.53}_{1.35} \times 10 ^{-27} \quad \textrm{\cite{J}}. 
\end{equation}

\begin{table}
\caption{Reported value for Hubble constant\cite{Chen}}\label{tab1}
\centering
\scalebox{0.9}{
\begin{tabular}{ |p{4.8cm}|p{1.5cm}||}
\hline

Gravitational lens system & $H$ \\ \hline
\hline
HE0435-1223&   $82.8^{+9.4}_{-8.3}$   \\ \hline
PG1115+080&  $70.1^{+4.5}_{-5.3}$ \\ \hline
RXJ1131-1231&   $77.0^{+4.0}_{-4.6}$ \\ \hline
The joint adaptive optics(AO) results&   $75.6^{+3.2}_{-3.3}$ \\ \hline
The joint adaptive optics(AO) + $HST$ results&   $76.8^{+2.6}_{-2.6}$ \\ \hline
\hline
\end{tabular}
}
\end{table}

\section{Local Group}

Mass = $2.3 \times 10^{12} M_{\odot}$ \cite{LG}, Radius = $1.5$ $Mpc$, $r_{crit}= 1.46$ $Mpc$.

Applying Eqs.(4),(6) to the observer at the center of the Local Group (LG) the galaxies in a certain vicinity of the LG will be repelled and it will cause a local H-flow; see also \cite{Kar,Ch}. Namely, we can define the distance  at which the flow will be exactly equal to the reported value (in $Mpc$)

\begin{equation}\label{rLG}
2.010<r< 2.137\, \quad \textrm{\cite{M106}};  
\end{equation}
\begin{equation}
2.007<r< 2.168\ \quad \textrm{\cite{AR1}};
\end{equation}
\begin{equation}
2.000<r< 2.147\, \quad \textrm{{\cite{AR2}}};
\end{equation} 
\begin{equation}
2.002<r< 2.153\, \quad \textrm{\cite{AR3}};
\end{equation}
\begin{equation}
1.802<r< 2.218\, \quad \textrm{\cite{J}}.  
\end{equation}

\section{Virgo Cluster}

Mass = $1.2 \times 10^{15} M_{\odot}$ \cite{VC}, Radius = $2.2$ $Mpc$, $r_{crit}= 11.80$ $Mpc$.

In this case, it should be noticed that although $r_{crit}$ is outside the cluster, it is smaller than the distance between the cluster and LG. This is due to the fact, that the center of the cluster is located in $16.5$ $Mpc$ from us. Again, one can obtain the distance on which a gravitational system centered in the center of Virgo cluster can cause a H-flow (in $Mpc$)

\begin{equation}\label{rVC}
16.18<r< 17.20\, \quad \textrm{\cite{M106}}; 
\end{equation}
\begin{equation}
16.16<r< 17.45\, \quad \textrm{\cite{AR1}};
\end{equation}
\begin{equation}
16.10<r< 17.29 \, \quad  \textrm{\cite{AR2}};
\end{equation}
\begin{equation}
16.11<r< 17.33\ , \quad \textrm{\cite{AR3}};
\end{equation} 
\begin{equation}
14.51 < r < 17.72 \quad \textrm{\cite{J}}.
\end{equation}
A remarkable consequence of this result is the following. By comparing the limits of the above relation with the distance between Virgo cluster and LG, one can state that the gravitational repulsion produced by Virgo cluster can repel the whole LG in an exact accordance with the reported value of $H$. Inversely, we, as the observers located within LG, can observe the Virgo cluster as moving away from us exactly according to local H-flow.

A new estimation for the virial mass of Virgo cluster is obtained in \cite{Vir}, equal to $(6.3 \pm 0.9) \times 10^{14}$ $M_{\odot}$. Within the $\Lambda$-gravity, from the appeared additional ``effective mass" $\frac{\Lambda c^2 r^3}{3G}$ we get the following upper limit for $\Lambda$
\begin{equation}\label{virL}
\Lambda \leq \frac{3 G \mathrm{E}(M_{vir})}{c^2 r^3}= 4.05 \times 10^{-51} \quad m^{-2}.
\end{equation} 

On the other hand, the virial parameters also can be used (with obvious precautions regarding the degree of virialization) to find an upper limit for $\Lambda$, namely,
\begin{equation}\label{vir}
( 1- \frac{\mathrm{E}(\sigma)}{\sigma})^2 \leq 1- \frac{\Lambda c^2 r^3}{3 G M_{vir}}= 6.52 \times 10^{-52}\,\, m^{-2}.
\end{equation}
Note that, although these limits are considerably small, neither contradicts the reported value for $\Lambda$ of {\it Planck} \cite{Pl}.

\section{Virgo Supercluster}
Mass = $1.48 \times 10^{15} M_{\odot}$ \cite{VS}, Radius = $16.5$ $Mpc$, $r_{crit}= 12.66$ $Mpc$.

This case means that one has to consider both LG and Virgo cluster within a larger scale structure, i.e. one gets ($Mpc$)  
\begin{equation}\label{rVS}
17.35<r< 18.45\, \quad \textrm{\cite{M106}}; 
\end{equation}
\begin{equation}
17.33<r< 18.72\, \quad \textrm{\cite{AR1}};
\end{equation}
\begin{equation}
17.27<r< 18.54 \,  \quad \textrm{\cite{AR2}};
\end{equation}  
\begin{equation}
17.28<r< 18.59\,  \quad \textrm{\cite{AR3}};
\end{equation} 
\begin{equation}
15.56 < r < 19.00 \quad \textrm{\cite{J}}.
\end{equation}
Note that, while the centers of Virgo Supercluster and Virgo cluster are considered to be identical, the LG is in the outskirts of the supercluster. Then, the center where the local H-flow caused by Virgo Supercluster repels the LG is located at $r= 1.30$ $Mpc$ away from the center of the LG. In this sense, by comparing the reported data of LG, it turns out that this radius is smaller than the radius of LG and even smaller than $r_{crit}$, which can be regarded as an indicator of gravitational boundness for a system.

\section{Laniakea Supercluster}

The mass of the revealed Laniakea supercluster \cite{L} is evaluated of the order of $10^{17}$ $M_\odot$. Here, we can perform an inverse analysis, to obtain an estimation of the mass according to Eq.(\ref{rcrit}) and Eq.(\ref{Hl}). Note that, although theses two equations seem similar, the nature of their analyses is totally different. Namely, Eq.(\ref{rcrit}) is obtained via a dynamical analysis, while Eq.(\ref{Hl}) is related to the gravitational energy. In this sense, by taking $M = 10^{17}$ $M_\odot$, $r_{crit}$ will be 51 $Mpc$. Now, considering the reported radius of Laniakea and the fact that it is located in 77 $Mpc$ from us, we can use Eq.(\ref{Hl}) to obtain the limits for the mass.  Namely, considering the reported values of $H$ \cite{M106,AR1,AR2,AR3,J}, one can find an (average) estimation
\begin{equation}\label{ML}
1.03 \times 10^{17} < M / M_\odot< 1.38 \times 10^{17}.
\end{equation}
The main consequence of this result is that, by considering the reported local value of $H$ and Eq.(\ref{Hl}), we are able to have a mass estimation which agrees with the reported data.

Thus, the Eqs.(\ref{mod}) and (\ref{FandU}), via Eq.(\ref{Hl}) and (\ref{Hg}), enable to define {\it local} and {\it global} Hubble constants and, hence, self-consistently - without any free parameter - describe the  observational data on the galaxy distribution and their flows in the vicinity of the Local Group and Virgo supercluster.

It should be stressed that, the obtained limit in Eq.(\ref{rel}) and the corresponding mass and radius for all the above analyzed cases, guarantee that we are in the weak-field limit regime. Thus, we are justified to analyze the dynamics of the said objects based on the weak-field of $\Lambda$-gravity and compare them with the results of \cite{M106,AR1,AR2,AR3,J}. The results of our analysis shows that for all above (hierarchical) systems, we will have the exact amount of density which can cause the recession of objects according to the measured H flow in \cite{M106,AR1,AR2,AR3,J}. Namely, the distance where the gravitational repulsion of the central object according to $\Lambda$-gravity forces the galaxies to move away is in complete agreement with the reported value of local H and its corresponding density.

 Thus, for the considered hierarchical systems of different scales, the galaxies move away from their centers starting from a critical distance, in accord to the presence of $\Lambda$ in Eq.(\ref{mod}). This is what we call the "local H-flow". Accordingly, in order to show the difference of this flow with the relativistic flow of objects at cosmic scales, we have obtained for all systems, the relevant distance at which the objects start to flow. We have found that for all five reported values of $H$, the corresponding distance is in agreement with the observations.

\section{Probes by galactic dark matter}

The weak-field GR given by Eqs.(\ref{mod}) and (\ref{FandU}) has been applied to describe the dynamics of galactic halos, galaxy groups and clusters \cite{G,GS4} by means of the virial theorem for the gravitational potential containing besides the Newtonian term also the one with the cosmological constant $\Lambda$. So, as in \cite{GS4} at comparison with observational data, the current numerical value of the cosmological constant $\Lambda$ has to be smaller than the error of velocity dispersion. 
We will now extend such an analysis to two categories of extremal cases i.e. to galaxies with no DM and galaxies made up of DM only. In all these cases, the reported data are in accordance with Newtonian dynamics. Namely, the measured velocity dispersion $\sigma$ is related to dynamical mass $M_{dyn}$ via the following relation
\begin{equation}\label{Newton}
\sigma^2 = \frac{G M_{dyn}}{R},
\end{equation}
where $R$ is the typical radius of galaxy. Clearly, the above dynamical equation is obtained by considering the Newtonian gravity. However, by replacing the Newtonian gravitational force with $\Lambda$-gravity according to Eq.(\ref{mod}) we will get
\begin{equation}\label{Lambda}
\sigma^2 = \frac{G M_{dyn}}{R} - \frac{\Lambda c^2 R^2}{3}.
\end{equation}
Comparing Eqs.(\ref{Newton}, \ref{Lambda}), it turns out that observed value of $\sigma$ in the context of $\Lambda$-gravity should be smaller than Newtonian case. Thus, in order to be a self consistent theory, the theoretically obtained value of $\sigma$ in the context $\Lambda$-gravity should be larger than observed value of $\sigma$ which is based on Newtonian gravity. Consequently, as a method to check the validity of $\Lambda$-gravity theory, we can find the upper limit for the numerical value of $\Lambda$ as follows
\begin{equation}\label{Error}
(\frac{\sigma - \mathrm{E} (\sigma)}{\sigma})^2 \leq 1- \Lambda \frac{c^2 R^3}{3 G M_{dyn}},
\end{equation}
where $\mathrm{E} (\sigma)$ is the error limit of velocity dispersion reported by observations. In this sense, it is expected that the obtained upper limits for $\Lambda$ must be larger than numerical value of $\Lambda = 1.1 \times 10^{-52}$ $m^{-2}$ which have been reported by \textit{Planck} satellite.

Analysis of such extreme cases can pose constraints over various theories of gravity and even rule them out \cite{T}.  

Indeed, by assuming Newtonian dynamics, the recent study \cite{DF2} proposes two estimations for dynamical mass of NGC 1052-DF2. Consequently, the upper limit over the $\Lambda$, for intrinsic and nominal velocity dispersions i.e. $\sigma_{int}$ and $\sigma_{DF2\star}$ will be
\begin{equation}\label{DF2}
\textrm{$\sigma_{int}$}: \quad \Lambda \leq 3.41 \times 10^{-49}, \qquad \textrm{$\sigma_{DF2\star}$}: \quad \Lambda \leq 5.47 \times 10^{-49}.
\end{equation}
For the second DM-missing galaxy NGC 1052-DF4, we also obtain the upper limits as follows:
\begin{equation}\label{DF4}
\textrm{$\sigma_{int}$}: \quad \Lambda \leq 1.37 \times 10^{-50}, \quad \quad \textrm{$\sigma_{stars}$}: \quad \Lambda \leq 1.85 \times 10^{-50},
\end{equation}
where $\sigma_{stars}$ refers to velocity dispersion obtained by considering the stars alone.

For the other extreme category we check the structure of Dragonfly 44 as one of best known ultra diffuse galaxies (UDG) \cite{Dr44}. Here by considering the total dynamical mass $M_{dyn}$ within the half-light radius i.e. $r = 4.3$ $kpc$ equal to $0.7^{+0.3}_{-0.2} \times 10^{10} M_{\odot}$ we  have
\begin{equation}
\Lambda \leq 2.61 \times 10^{-48}.
\end{equation}

Thus by considering the results of both categories of objects - galaxies lacking DM and the one made almost entirely of DM - it turns out that the modification of gravity according to Eq.(\ref{mod}) not only is able to describe these structures, but fits the considered weak-field GR with the numerical value of $\Lambda$ not contradicting the observational data on these extremal astrophysical structures.
 
Besides the above two categories of galaxies, a new group denoted as \textit{DM deficient dwarf galaxies} has been studied in \cite{DMDG}. For them it has been reported that the matter content consists mainly of baryons. We start our discussion by checking the velocity of galaxies according to Eq.(\ref{mod}) i.e.
\begin{equation}\label{CV}
V^2_{cir} = \frac{GM_{dyn}}{r} - \frac{\Lambda c^2 r^2}{3},
\end{equation}
where $M_{dyn}$ is the total dynamical mass. Thus, by taking the reported values of these galaxies we find the error limits of $\Lambda$. The results are shown in Table \ref{tab2}.  The $w20$ denotes the $20 \%$ of the HI line width which has been considered as indicator of the gas velocity. Considering the results of Table \ref{tab2}, it becomes clear that again there is no contradiction between $\Lambda$-modified gravity and the observed parameters of the galaxies. 

\begin{table*}
\caption{Constraints on $\Lambda$ for DM deficient dwarf galaxies}\label{tab2}
\centering
\scalebox{0.6}{
\begin{tabular}{ |p{2cm}|p{2.4cm}|p{1.2cm}|p{1.2cm}||p{1.8cm}|| }
\hline

Galaxy & $\log M_\mathrm{dyn} (M{}_{\odot})$ & $w20$ (km/s) & $w20_{er}$ (km/s) & $\Lambda$ $(m^{-2})$ $\leq $  \\ \hline
\hline
AGC 6438&   9.444 & 80.36 & 2.03 & $9.57 \times 10^{-50}$ \\ \hline
AGC 6980&  9.592 & 56.63 & 1.54 & $6.37 \times 10^{-51}$ \\ \hline
AGC 7817&   9.061 & 82.37 & 4.45 & $1.36 \times 10^{-48}$ \\ \hline
AGC 7920&   8.981 & 79.03 & 2.6 & $9.47 \times 10^{-49}$ \\ \hline
AGC 7983&   9.046 & 46.12 & 0.83 & $1.52 \times 10^{-50}$ \\ \hline
AGC 9500&   9.092 & 39.08 & 0.31 & $2.02 \times 10^{-51}$ \\ \hline
AGC 191707&  9.08 & 49.27 & 1.21 & $2.64 \times 10^{-50}$ \\ \hline
AGC 205215&  9.706 & 72.5 & 4.41 & $3.65 \times 10^{-50}$ \\ \hline
AGC 213086&  9.8 & 78.35 & 4.33 & $3.43 \times 10^{-50}$ \\ \hline
AGC 220901&  8.864 & 45.38 & 0.74 & $2.91 \times 10^{-50}$ \\ \hline
AGC 241266&  9.547 & 52.82 & 1.98 & $7.08 \times 10^{-51}$ \\ \hline
AGC 242440&  9.467 & 42.47 & 1.18 & $2.06 \times 10^{-51}$ \\ \hline
AGC 258421&   10.124 & 87.79 & 8.53 & $2.63 \times 10^{-50}$ \\ \hline
AGC 321435&   9.204 & 56.83 & 4.41 & $1.08 \times 10^{-49}$ \\ \hline
AGC 331776&   8.503 & 29.59 & 2.9 & $6.79 \times 10^{-50}$ \\ \hline
AGC 733302&   9.042 & 48.36 & 0.99 & $2.37 \times 10^{-50}$ \\ \hline
AGC 749244&   9.778 & 70.87 & 4.91 & $2.59 \times 10^{-50}$ \\ \hline
AGC 749445&   9.264 & 54.51 & 3.06 & $4.67 \times 10^{-50}$ \\ \hline
AGC 749457&   9.445 & 58.68 & 5.49 & $5.16 \times 10^{-50}$ \\ \hline
\hline
\end{tabular}
}
\end{table*}

In addition to above extreme cases  62 dwarf spheroidals (dSphs) in the Local Group (LG) are considered as another sample to analyze the validity of different modified theories of gravity and the paradigm of DM. Namely, the study of dSphs surrounding the Milky Way has suggested those are DM-free structures \cite{SpMW}. Here, by considering Eq.(\ref{Error}) we have obtained error limits of $\Lambda$ for them. The results are exhibited in Table \ref{tab3}.

\begin{table*}
\caption{Constraints on $\Lambda$ for 24 dwarf galaxies surrounding the Milky Way} \label{tab3}
\centering
\scalebox{0.6}{
\begin{tabular}{ |p{1.8cm}|p{1.8cm}|p{2.1cm}||p{1.8cm}|| }
\hline

Galaxy & $\sigma (km/s^{2})$ & $r$ (pc) & $\Lambda$ $(m^{-2})$ $\leq $ \\ \hline
\hline
Aquarius2& $5.4\pm 3.4$ & $160.0 \pm 24.0$ & $1.32 \times 10^{-46}$ \\ \hline
Bootes1&  $2.4\pm 0.9$ & $192.5\pm 5.039$ & $9.72 \times 10^{-48}$ \\ \hline
Carina&   $6.6\pm 1.2$ & $303.1\pm 2.952$ & $1.32 \times 10^{-47}$ \\ \hline
Coma& $4.6\pm 0.8$ & $68.59\pm 3.615$ & $1.18 \times 10^{-46}$ \\ \hline
CraterII&   $2.7\pm 0.3$ & $1066\pm 86$ & $1.05 \times 10^{-49}$ \\ \hline
CVenI&   $7.6\pm 0.4$ & $437.9\pm 12.59$ & $2.28 \times 10^{-48}$ \\ \hline
CVenII&  $4.6\pm 1.0$ & $70.83 \pm 11.22$ & $1.42 \times 10^{-46}$ \\ \hline
Draco&  $9.1\pm 1.2$ & $222.4\pm 2.079$ & $3.30 \times 10^{-47}$ \\ \hline
Draco2&  $2.9\pm 2.1$ & $20.73 \pm 7.639$ & $2.71 \times 10^{-45}$\\ \hline
Fornax&   $11.7\pm 0.9$ & $792.5 \pm 2.837$ & $2.44 \times 10^{-48}$ \\ \hline
Hercules&   $3.7\pm 0.9$ & $221.1\pm 17.4$ & $1.07 \times 10^{-47}$ \\ \hline
LeoI&   $9.2\pm 1.4$ & $287.9 \pm 2.133$ & $2.35 \times 10^{-47}$\\ \hline
LeoII&   $6.6\pm 0.7$ & $164.7 \pm 1.926$ & $2.52 \times 10^{-47}$\\ \hline
LeoIV& $3.3 \pm 1.7$ & $114.3 \pm 12.03$ &  $7.59 \times 10^{-47}$ \\ \hline
LeoV&  $2.3\pm 3.2$ & $50.41\pm 16.15$ & $6.90 \times 10^{-46}$ \\ \hline
Sagittarius&  $11.4 \pm 0.7$ & $1636.0 \pm 52.78$ & $4.31 \times 10^{-49}$ \\ \hline
Sculptor&  $9.2\pm 1.4$ & $276.4 \pm 0.9872$ & $2.54 \times 10^{-47}$ \\ \hline
Segue1&  $3.9 \pm 0.8$ & $ 24.11 \pm 2.79$ & $8.31 \times 10^{-46}$ \\ \hline
Sextans&   $7.9 \pm 1.3$& $412.1\pm 2.993$ & $9.19 \times 10^{-48}$ \\ \hline
TucanaII &  $8.6 \pm 3.5$ & $156.3\pm 23.68$ & $2.08 \times 10^{-46}$ \\ \hline
UMaI&  $7.6\pm 1.0$ & $234.2 \pm 10.01$ & $2.07 \times 10^{-47}$ \\ \hline
UMaII& $6.7 \pm 1.4$ & $ 136.3 \pm 5.325$ & $7.83 \times 10^{-47}$ \\ \hline
UMi& $9.5\pm 1.2$ & $407.0 \pm 2.0$ & $1.02 \times 10^{-47}$ \\ \hline
Willman1& $4.3\pm 2.3$ & $27.7\pm 2.4$ & $2.29 \times 10^{-45}$ \\ \hline
\hline
\end{tabular}
}
\end{table*}

Moreover, considering Eq.(\ref{mod}), the radial acceleration will be written as
\begin{equation}\label{aL}
a(r) = \frac{G M_{tot}}{r^2} - \frac{\Lambda c^2 r}{3},
\end{equation}
where $M_{tot}$ is the total mass (both ordinary and DM) of the configuration according to \cite{Sp}. Consequently, the constrains over $\Lambda$ will be obtained. For 20 of them these limits are shown in Table \ref{tab4}.

\begin{table*}
\caption{Constraints on $\Lambda$ for 20 dwarf spheroidals of LG} \label{tab4}
\centering
\scalebox{0.6}{
\begin{tabular}{ |p{2.7cm}|p{2.4cm}|p{1.2cm}||p{1.8cm}|| }
\hline

Galaxy & $\log a(r) (m/s^{2})$ & $r$ (pc) & $\Lambda$ $(m^{-2})$ $\leq $ \\ \hline
\hline
Bootes I&  -$11.14\pm 0.15$ & $283\pm 7$ & $8.09 \times 10^{-48}$\\ \hline
Bootes II&  -$9.75 \pm 0.63$ & $61 \pm 24$ & $2.41 \times 10^{-45}$ \\ \hline
Canes Venatici I& -$11.06\pm 0.05$ & $647 \pm 27$ & $1.58 \times 10^{-48}$ \\ \hline
Canes Venatici II& -$10.69\pm 0.19$ & $101\pm 5$ & $7.75 \times 10^{-47}$ \\ \hline
Carina& -$10.81\pm 0.18$ & $273\pm 45$ & $2.08 \times 10^{-47}$ \\ \hline
Coma Berenices &   -$10.59 \pm 0.16$ & $79\pm 6$ & $1.08 \times 10^{-46}$ \\ \hline
Draco&  -$10.48\pm 0.12$ & $244\pm 9$ & $3.54 \times 10^{-47}$ \\ \hline
Fornax& -$10.77\pm 0.08$ & $792 \pm 58$ & $3.90 \times 10^{-48}$ \\ \hline
Hercules&  -$11.11\pm 0.22$ & $175\pm 22$ & $1.90 \times 10^{-47}$ \\ \hline
Hydra II&   -$10.65 \pm 0.12$& $88\pm 17$ & $6.64 \times 10^{-47}$ \\ \hline
Leo I&  -$10.56\pm 0.06$ & $298 \pm 29$ & $1.29 \times 10^{-47}$ \\ \hline
Leo II& -$10.71\pm 0.14$ & $219 \pm 52$ & $2.65 \times 10^{-47}$ \\ \hline
Leo IV&   -$11.15\pm 0.47$ & $149\pm 47$ & $3.39 \times 10^{-47}$ \\ \hline
Leo V&   -$11.35\pm 0.88$ & $125\pm 47$ & $3.35 \times 10^{-47}$ \\ \hline
Leo T&   -$10.47\pm 0.19$ & $160 \pm 10$ & $8.12 \times 10^{-47}$ \\ \hline
Sculptor& -$10.58 \pm 0.13$ & $311 \pm 46$ &  $2.36 \times 10^{-47}$ \\ \hline
Sextans&  -$11.09 \pm 0.15$ & $748 \pm 66$ & $3.43 \times 10^{-48}$ \\ \hline
Ursa Minor&  -$10.66\pm 0.12$ & $398\pm 44$ & $1.43 \times 10^{-47}$ \\ \hline
Ursa Major I&   -$11.35\pm 0.88$ & $125\pm 47$ & $3.35 \times 10^{-47}$ \\ \hline
Ursa Major II&   -$10.47\pm 0.19$ & $160 \pm 10$ & $8.12 \times 10^{-47}$\\ \hline
\hline
\end{tabular}
}
\end{table*}

It is worth noticing that although several parameters are taken into consideration in actual observations, they are based on fundamental relations governing the dynamics of objects according to Newtonian gravity. What we have done in this section is to modify the underlying dynamical equations in the context of $\Lambda$-gravity (i.e. Eqs.(\ref{CV}),(\ref{aL}) and analyze the error limits accordingly assuming that all the observational complications are the same.

Thus, the galaxy samples which were previously used to test and/or reject certain dark matter models, here are shown to be in full compatibility with the $\Lambda$-gravity predictions.

\section{Conclusions}

The Hubble tension problem, now attracting much attention, is shown to be resolvable conceptually and quantitatively by the $\Lambda$-gravity, as modified weak-field General Relativity \cite{G,GS1}.

We show that the suggested approach defines a ladder of distance scales for galaxy distribution hierarchy, from the Local group to the Virgo and Laniakea superclusters, which links their local dynamics to the cosmological parameters.   

For those considered hierarchical systems of different scales we conclude that the galaxies move away from their centers starting from a critical distance, in accord to the presence of $\Lambda$ in Eq.(\ref{mod}) i.e. contribute to the "local H-flow". We show the difference of this flow with the relativistic flow at cosmological scales, i.e. we obtain for all systems the relevant distances at which the objects would start to participate the {\it global} flow and show their agreement with the observations.

Importantly, the $\Lambda$-gravity is also shown to agree with the data on {\it extremal} galaxies, i.e. those claimed as dark matter rich galaxies and no dark matter ones. Several independent galaxy samples are considered, all shown with data compatible to modified gravity constraints. Again, the principal point in our analysis and in comparison to the observational data is the absence of any additional or free theoretical parameters.

Future more refined observational surveys of galaxy distribution and dynamics in the vicinity of the Local group and the Virgo supercluster can be decisive in testing the modified weak-field General Relativity, with direct impact on the nature of the dark sector.

\end{document}